\documentstyle[12pt]{article}
\def\beq{\begin{equation}}
\def\eeq{\end{equation}}
\def\ben{\begin{eqnarray}}
\def\een{\end{eqnarray}}
\def\bea{\begin{array}}
\def\eea{\end{array}}

\begin{document}

\baselineskip=20pt


\begin{flushright}
Prairie View A \& M, HEP-12-95\\
December 1995 \\
Los Alamos National Lab bulletin board: hep-ph/9603418


\end{flushright}

\vskip.15in

\begin{center}
{\Large\bf  CP Symmetry and Fermion Masses in\\
   $O(10)$ Grand Unification Models
 }

\vskip .3in

{\bf Dan-Di Wu}\footnote{ $~$E-mail:  wu@hp75.pvamu.edu or 
danwu@physics.rice.edu }

{\sl HEP, Prairie View A\&M University, Prairie View, TX 77446-0355, USA}

and 

{\bf Yue-Liang Wu}\footnote{ ~E-mail: ylwu@mps.ohio-state.edu}

{\sl Department of Physics, Ohio State University, Columbus, OH 43210 }

\vskip 2.5in

\end{center}
 {\small O(10) grand unification models which do not necessarily
have an extra global symmetry are discussed, taking the model with one 
10-plet in the Yukawa sector as an example. 
A strong correlation between mass ratios and CP is found.  
The mass relation 
 $m_t/m_b=v_u/v_d$ is recovered when $G_W=0$; and  another special relation
$m_t/m_b=G_E/G_W$ appears when $v_d=0$, where $G_{E,W}$ are Yukawa coupling
constants and $v_{u,d}$ are VEVs. To facilitate this 
discussion, a set of $O(10)$ $\gamma$-  
 matrices  is offered based on a physical representation of the 
 spinors and that of the vector of the $SO(10)$ group.
Flavor changing neutral currents in such models  are also 
discussed.}
\clearpage

{\bf \large  I. Motivations }
\vskip.1in
The $SO(10)$ group is one of the most prominent groups for unification 
theories[1]. 
In most previous studies of $SO(10)$ grand unification theories,
extra global symmetries are assumed either explicitly or implicitly.
While these extra global symmetries can be  justified from 
supersymmetry or superstring theories, a pure $SO(10)$ grand unification theory
may still be of interest.
  For example, the flavor 
changing neutral currents (FCNC) in a $SO(10)$ 
model 
 is somewhat subtle. Taking the model with one 10-plet mass provider 
(minimal O-ten models,  MOTM) as an example, the 10-plet is 
decomposed into $SU(5)$ representations as
$10=5_1+5_2^*$.
 The fermions 
16 and 16$^*$ are decomposed as
$16=10+5^*+1,$ and $ 16^*=10^*+5+1$.
Therefore there are two sets of possible mass terms:
$$5_1 \ 10^*\ 5,\hskip.3in 5^*_2\ 10^*\ 10^*;$$
and
$$5_1\ 10\ 10,\hskip.3in 5^*_2\ 10\ 5^*.$$
There will be no FCNC at 
low energies  for MOTM with three generations of fermions, 
if only one of the two sets contributes. On the other hand,
if only one of the 5-plets contribute, but both 
couple to fermions, as they must do because they are in the same multiplet,
there can still be FCNC. 
\vskip.1in
Indeed, one will see that the top-bottom mass ratio can be written as
\beq
|\frac{m_t}{m_b}|=\left\{\frac{[Re(g_ev_1+g_ov_6)]^2+[Im(g_ev_6+g_ov_1)]^2}
{ [Re(g_ev_1-g_ov_6)]^2+[Im(g_ev_6-g_ov_1)]^2}\right\}^{\frac{1}{2}}. 
\eeq
Here, 
$
v_1=(v_u+v_d)/2,\hskip.15in v_6=(v_u-v_d)/2
$
where $v_u$ and $v_d$ are  the VEVs of $5_1$ and $5^*_2$ 
respectively of the 10-plet Higgs. $g_e$ and $g_o$ are respectively coupling 
constants of $O(10)$
parity even and odd  terms. The 10-d space reflection 
 is not an element of the $SO(10)$ group. 
Therefore
$O(10)$ is also of interest.
 One can see from this formula that
in order to have $|m_t/m_b|\ne 1$, not only one needs both vacua $v_1$ 
and $v_6$ but also both couplings $g_e$ and $g_o$. 
 When there is a maximal CP mixing, 
$g_o=\pm ig_e$, the mass ratios will be adversely 
affected in the MOTM. 
\vskip.1in
Two special cases are  worth noting: 1) $v_1=v_6$: This means that there is
only one nonzero VEV, $v_u\ne 0,\ 
v_d=0$, which is typical for a vector (the 10-plet) to develop VEV.
  In this case, one obtains $m_t/m_b=(g_e+g_o)/(g_e-g_o)$.
 2) The Yukawa couplings satisfy the condition for
being self-dual, $g_o=g_e$.  The definition of dual (denoted by E for 
convenience) and anti-dual (denoted by W)
 in $O(10)$ is similar to that of
left-handed (L) and right-handed (R) in the Lorentz group.  In this 
case one obtains a two Higgs doublet model with either supersymmetry
or a global $U(1)$ symmetry[2]
at low energies. 
\vskip.1in
In general, the counterpart of a MOTM 
at low energies is a general two Higgs doublet model[3] with FCNC 
and a complicated
relation between $m_t/m_b$ and $v_u/v_d$. 
The second special case, particularly when it is resulted
from supersymmetry, is widely applied  
for discussions of $SO(10)$ mass relations[4]. In this note we will analyze
the $O(10)$ models without any constraints.   

\vskip.1in
Therefore a one 10-plet Higgs $O(10)$  model (MOTM) 
may in general correspond to a two
 Higgs doublet model with FCNC. If more Higgs multiplets are involved in the 
Yukawa sector, then there will be more FCNCs. In either case,  
  the Yukawa coupling constants can be complex,
 which may cause explicit CP violation. 
In addition to this, spontaneous CP violation 
due to a relative phase of the VEVs may appear 
in all the electro-weak interactions.  
In general, one may need eight  U- matrices in order to 
diagonalize the mass matrices of the up-, down-, neutrino- and lepton-
mass matrices:
$
U^U_L, U^U_R;\,\,U^D_L,U^D_R;\,\ U^\nu_L,U^\nu_R;\ \ 
{\rm and }\ \ U^l_L,U^l_R.
$
All of them can be physically relevant; in other words, in addition to
the CKM matrix 
$
V_{CKM}= U^U_LU^{D\dagger}_L
$;
 the matrix
$
V^\prime=U^U_RU^{D\dagger}_R
$
appears in the right-handed charged current gauge interactions; the matrix
$
\tilde G^{U} =U^U_LG^U_YU^U_R
$
leads to the  scalar mediated FCNC interactions 
 among up-type quarks, where $G_Y^U$ is the matrix 
 of Yukawa couplings of this 
scalar. Unless $G_Y^U$ of 
a Higgs doublet is proportional to the corresponding mass matrix, 
FCNC mediated by this Higgs field is in general nontrivial.
CP violation can in principle appear in any of above   interactions.  
$V^\prime$  and $\tilde G^U$ and 
 the alike represent physics beyond the CKM matrix[3, 5]. 
\vskip.1in

This work is  devoted to the general relations among mass ratios, FCNC,
and CP violation in an arbitrary $O(10)$ model. The MOTM will be taken as an 
explicit example. However,  
the results can be applied to any non-minimal $O(10)$ models. Before 
such a discussion,  a  set of  explicit 
$\gamma$- matrices will be provided in Section II. The properties of
 mass operators will be 
discussed in detail in Section III. The special cases will be reviewed
in terms of the dual (E) and anti-dual (W) coupling constants. In Section IV,
symmetries beyond $O(10)$ are discussed which may help to forbid the W term
or the E term.
 
\vskip.1in

{\bf \large  II.  The $SO(10)$ Gamma Matrices and Mass Operators}
\vskip.15in

There have been discussions on the group of $SO(10)$, since it was 
recognized as a potential candidate group for  grand unification theories 
(GUT)[1, 6]. A different approach will be taken in this work. 
The components of
the fundamental spinor and vector representations will be assigned first, in
terms of the familiar quantum numbers, such as color, flavor, and $B-L$.
Then the $\gamma$- matrices will be built up on this specific basis. 

\vskip.1in
 A fermion field can be seen as a sum of the left-handed and the 
right-handed parts
$
\psi=\psi_L+\psi_R, \ \ \ \psi_{R,L}=\frac{1}{2}(1\pm\gamma_5)\psi.
$
A generic mass term is 
\beq
\bar\psi_L\psi_R=\psi^{cT}_LC\psi_R=\psi_R^TC\psi_L^c.
\eeq
where $\psi^c=C\bar\psi^T $, and $C=i\gamma_{_2}\gamma_{_0}$ is the C-matrix 
of the Lorentz group, $C^T=C^{-1}=-C$. 
From here on, all Lorentz group matrices will be underlined, in order to 
distinguish them from the $SO(10)$ matrices.
\vskip.1in

The present task is to find ten $32\times 32$  matrices which satisfy the 
Clifford algebra: 
\beq
\left\{\gamma_{_M},\gamma_{_N}\right\}=\gamma_{_M}\gamma_{_N}+
\gamma_{_N}\gamma_{_M}=2\delta_{_{MN}}I^{32}.
\eeq
The anti-commuting relation (3) keeps its validity under 
orthogonal transformations
$
\gamma_{_M}^\prime=a_{_{MN}}\gamma_{_N}.
$
 In addition one  has a  freedom to 
choose the components of the reducible spinor $2^5=16+16^*$.  
\vskip.1in
 The two irreducible spinors of $SO(10)$ are 
 represented here by
$\psi $ and $\psi^c$.  They are used to represent respectively Lorentzan
left-handed and right-handed Weyl fields in one family of fermions
\beq
\psi^T= (u_L,d_L,d^c_R,u^c_R,\nu_L,e_L,e^c_R,\nu^c_R),
\eeq
and the $16^*$- plet is just its charge conjugate.
The color indices (from 1 to 3) for the quarks are suppressed.
The arrangement of the components in (4) 
complies with a $SU(3)\times SU(2)_L\times SU(2)_R\times U(1)$ 
decomposition of the
$SO(10)$ spinor: 
$
16=(3,2,1,+1/3)+(3^*,1,2,-1/3)+(1,2,1,-1)+(1,1,2,+1),
$
where the fourth number in each parenthesis is the $B-L$ 
quantum number\footnote
{The $SU(5)\times U(1)$ decomposition can be reached by rearranging the 
components to the following form:\\
 $\psi^T=
(u_L,d_L,u^c_R,e^c_R,d^c_R,\nu_L,e_L,\nu^c_R)$.}.
The 32-d reduced representation is chosen as $\Psi$,
$
\Psi=\left(\bea{c}
\psi\\
\psi^c\eea\right).
$                                                                                                                                                                                                                           
\vskip.1in

It is convenient  to first define 10 
symmetric $16\times 16$ matrices in two groups:
$
\alpha_i,\ \beta_p, (i=1,2,3,4,5; p=6,7,8,9,10),
$
where $\alpha$s and $\beta$s mutually commute, 
while each groups make separate
Clifford  algebras, 
\beq
\{\alpha_{_i},\alpha_{_j}\}=2\delta_{_{ij}}I^{16},
\,\,\,\{\beta_p,\beta_q\}=2\delta_{_{pq}}I^{16};\,\,\,
\left[\alpha_{_i},\beta_p\right]=0.
\eeq
The gamma matrices are then simply
\ben
\bea{ccccc}
\gamma_{_I}&=&\left(\bea{cc}
0&\alpha_{_I}\\
\alpha_{_I}&0\eea\right),
&\,\,\, &(I=1,...5)\\
&&&&\\
\gamma_{_P}&=&\left(\bea{cc}
0&-i\beta_P\\
i\beta_P&0\eea\right).&\,\,\, &(P=6,...10)
\eea
\een

\vskip.1in
As mentioned before, the Clifford algebra is also invariant under orthogonal 
transformations, which changes the components of a 10-plet, for example.
Therefore, the specific form of gamma-matrices depends on how we choose
the components of a 10-plet. Using the fermion symbols, 
we represent quantum numbers of 
 the chosen basis components for 10-plet  as the following 
\footnote{The fermion symbols  used here are
for their  quantum numbers. For example, the quantum numbers of a term
$\nu_{_{1L}}\nu_{2R}^c$ are: electric charge $Q = 0$, lepton charge $L
= 0$, and 
$(T_{3L}, T_{3R})=(\frac{1}{2},-\frac{1}{2})$.
The attached subscripts (1 ro 2) are  useful to 
avoid  the 10-plet to be self-conjugated. 
Readers who do not prefer this basis for 10-plet, 
which mixes components with opposite quantum numbers, 
may read the next section for other representations in which no Clifford
algebra can be found though.} 
{\small
\ben\bea{c}
H=\frac{1}{2}\hskip5.3in\\
(
\nu_{_{1L}}\nu_{2R}^c+\nu^c_{2L}\nu_{_{1R}},
e_{_{1L}}\nu_{2R}^c+e^c_{2L}\nu_{_{1R}},
d^{1c}_{1R}\nu_{2R}^c+d^{1}_{2R}\nu_{_{1R}},
d^{2c}_{1R}\nu_{2R}^c+d^{2}_{2R}\nu_{_{1R}},
d^{3c}_{1R}\nu_{2R}^c+d^{3}_{2R}\nu_{_{1R}};\\
\\
i\nu_{_{1L}}\nu_{2R}^c-i\nu^c_{2L}\nu_{_{1R}},
ie_{_{1L}}\nu_{2R}^c-ie^c_{2L}\nu_{_{1R}},
id^{1c}_{1R}\nu_{2R}^c-id^{1}_{2R}\nu_{_{1R}},
id^{2c}_{1R}\nu_{2R}^c-id^{2}_{2R}\nu_{_{1R}},
id^{3c}_{1R}\nu_{2R}^c-id^{3}_{2R}\nu_{_{1R}}),\eea
\een}
the $\alpha$- and $\beta$- matrices  on the basis (7) are 
{\small
\ben
\bea{cc}
\alpha_{_1}=\left(\bea{cc|cc|cc}
&&&I^3&&\\
&&I^3&&&\\
\hline
&I^3&&&&\\
I^3&&&&&\\
\hline
&&&&&\tau_{_1}\\
&&&&\tau_{_1}&\eea\right),\,\,& \beta_6=\left(\bea{cc|cc|cc}
&&&-I^3&&\\
&&I^3&&&\\
\hline
&I^3&&&&\\
-I^3&&&&&\\
\hline
&&&&&\tau^\prime_2\\
&&&&-\tau^\prime_2&\eea\right);\\
&\\

\alpha_{_2}=\left(\bea{cc|cc|cc}
&&I^3&&&\\
&&&-I^3&&\\
\hline
I^3&&&&&\\
&-I^3&&&&\\
\hline
&&&&&\tau_3\\
&&&&\tau_3&\eea\right),
\,\,\,&
\beta_{7}=\left(\bea{cc|cc|cc}
&&-I^3&&&\\
&&&-I^3&&\\
\hline
-I^3&&&&&\\
&-I^3&&&&\\
\hline
&&&&&-I^2\\
&&&&-I^2&\eea\right);\\
&\\

\alpha_{_3}=\left(\bea{cc|cc|cc}
&\phi^3&&&B_1&\\
-\phi^3&&&&C_1&\\
\hline
&&&\phi^3&&B_1\\
&&-\phi^3&&&C_1\\
\hline
B^T_1&C^T_1&&&&\\
&&B_1^T&C_1^T&&\eea\right),\,\,&

\beta_8=\left(\bea{cc|cc|cc}
&-\phi^3&&&B_1&\\
\phi^3&&&&C_1&\\
\hline
&&&\phi^3&&-B_1\\
&&-\phi^3&&&-C_1\\
\hline
B^T_1&C^T_1&&&&\\
&&-B_1^T&-C_1^T&&\eea\right);\\
&\\

\alpha_{_4}=\left(\bea{cc|cc|cc}
&h^3&&&B_2&\\
-h^3&&&&C_2&\\
\hline
&&&h^3&&B_2\\
&&-h^3&&&C_2\\
\hline
B^T_2&C^T_2&&&&\\
&&B_2^T&C_2^T&&\eea\right),\,\,&

\beta_9=\left(\bea{cc|cc|cc}
&-h^3&&&B_2&\\
h^3&&&&C_2&\\
\hline
&&&h^3&&-B_2\\
&&-h^3&&&-C_2\\
\hline
B^T_2&C^T_2&&&&\\
&&-B_2^T&-C_2^T&&\eea\right);\\
&\\

\alpha_{_5}=\left(\bea{cc|cc|cc}
&\tilde\phi^3&&&B_3&\\
-\tilde\phi^3&&&&C_3&\\
\hline
&&&\tilde\phi^3&&B_3\\
&&-\tilde\phi^3&&&C_3\\
\hline
B^T_3&C^T_3&&&&\\
&&B_3^T&C_3^T&&\eea\right),\,\,
&
\beta_{10}=\left(\bea{cc|cc|cc}
&-\tilde\phi^3&&&B_3&\\
\tilde\phi^3&&&&C_3&\\
\hline
&&&\tilde\phi^3&&-B_3\\
&&-\tilde\phi^3&&&-C_3\\
\hline
B^T_3&C^T_3&&&&\\
&&-B_3^T&-C_3^T&&\eea\right),
\eea
\een
}
where
{\small
\ben
I^3=\left(\bea{ccc}
1&0&0\\
0&1&0\\
0&0&1\eea\right),\,\,\, h^3=\left(\bea{ccc}
0&0&-1\\
0&0&0\\
1&0&0\eea\right),\,\,\,\phi^3=\left(\bea{ccc}
0&0&0\\
0&0&1\\
0&-1&0\eea\right),\,\,\,\tilde\phi^3=\left(\bea{ccc}
0&1&0\\
-1&0&0\\
0&0&0\eea\right),
\een
}
and
{\small\ben
I^2=\left(\bea{cc}
1&0\\
0&1\eea\right),\,\,\,\tau_1=\left(\bea{cc}
0&1\\
1&0\eea\right),\,\,\,\tau^\prime_2=\left(\bea{cc}
0&-1\\
1&0\eea\right),\,\,\,\tau_3=\left(\bea{cc}
-1&0\\
0&1\eea\right),
\een
}
finally
{\small
\ben
\bea{c}
B_1=\left(\bea{cc}
0&1\\
0&0\\
0&0\\
\eea\right),\,\,\,
B_2=\left(\bea{cc}
0&0\\
0&1\\
0&0\\
\eea\right),\,\,\,
B_3=\left(\bea{cc}
0&0\\
0&0\\
0&1\\
\eea\right);\\
\\
C_1=\left(\bea{cc}
-1&0\\
0&0\\
0&0\\
\eea\right),\,\,\,
C_2=\left(\bea{cc}
0&0\\
-1&0\\
0&0\\
\eea\right),\,\,\,
C_3=\left(\bea{cc}
0&0\\
0&0\\
-1&0\\
\eea\right).
\eea
\een}
In all of these matrices, empty fields correspond to zeros. 
\vskip.08in
There are some additional matrices which will be useful for the
discussion of discrete symmetries.
First, $\gamma_{_{11}}$ is defined as the product 
of all the ten gamma matrices,
\beq
\gamma_{_{11}}=-i\gamma_{_1}\gamma_{_2}\cdot\cdot\cdot\gamma_{_{10}}=
{\rm diag}\left(I^{16},-I^{16}\right).
\eeq
Secondly the C-matrix is the product of the first five $\gamma$- matrices
\ben
C=\gamma_{_{1}}\gamma_{_{2}}\gamma_{_{3}}\gamma_{_{4}}
\gamma_{_{5}}=\left(\bea{cc}
0&I^{16}\\
I^{16}&0\eea\right).
\een 
$\gamma_{_{11}} $   can be used to construct  derived $\gamma$- matrices
$
\tilde \gamma_{_N}=\gamma_{_{11}}\gamma_{_N}, 
$
which satisfy the same conditions for a Clifford algebra, 
except for a sign
difference in normalization. 
\vskip.08in

To find out tensor representations decomposed from the product of 
two 32-spinor representations $\Psi_1$ 
and $\Psi_2$, let us first define
$
\bar\Psi=\Psi^{cT}=\Psi^T C.
$ 
The $a$-th order tensor $\gamma$- matrices are
\beq
\Gamma^{(a)}_{N_1\cdot\cdot\cdot N_a}=\frac{1}{a!} 
\gamma_{_{[N_1}}\cdot\cdot\cdot \gamma_{_{N_a]}},\ \ \ 
(a=0,\cdot\cdot\cdot, 2n).
\eeq
The bracket $[\cdot\cdot\cdot]$ for the subindices 
means 
anti-symmetrization. Obviously $\Gamma^{(10)}=i\gamma_{_{11}}$.
The  $SO(10)$ anti-symmetric tensor representations are simply
$\bar\Psi_1  \Gamma^{(a)}\Psi_2,$
where a ${\underline  C}$ of the Lorentz group 
is implied as is shown in Eq(2).
\vskip.08in

The second order $\gamma $- matrices $\Gamma^{(2)}$ are 
 the $SO(10)$ infinitesimal group operators 
 in the $32$ reduced 
representation. They are all anti-Hermitian and block diagonalized, 
therefore 
one has\footnote{However, when one of the ten dimensions 
(say, the 10th) is time-like, part of 
$\Gamma_{MN}$ are Hermitian due to the following definition: 
$\gamma_{_0}=i\gamma_{_{10}}$ which is anti- Hermitian. One then has
$$
e^{\Gamma^{(2)T}_{MN}a_{_{MN}}}C\gamma_{_0}
e^{\Gamma^{(2)}_{PQ}a_{_{PQ}}}=C\gamma_{_0}.
$$
This is the main difference between $SO(9,1)$ and $SO(10)$ groups.}
\beq
e^{\Gamma^{(2)T}_{MN}a_{_{MN}}}C\Gamma^{(0)}
e^{\Gamma^{(2)}_{PQ}a_{_{PQ}}}=C.
\eeq

5 among the 45 operators are diagonalized on the chosen basis. They are 
$
D_I=i\Gamma_{_{I,I+5}}=i\gamma_{_I}\gamma_{_{I+5}}.
$
Note that 
\ben
\bea{ccc}
\sqrt{\frac{1}{2}}\left({\it D}_2+{\it D}_1\right)
\hskip.15in&=&\sqrt{8}T_{3R},
\\

\sqrt{\frac{1}{2}}\left({\it D}_2-{\it D}_1\right)
\hskip.15in&=&\sqrt{8}T_{3L}, 
\\
                                                                                              
\sqrt{\frac{1}{3}}\left({\it D}_3+{\it D}_4+{\it D}_5\right)&=&-\sqrt{3}
\left(B-L\right),\\

\sqrt{\frac{1}{2}}\left({\it D}_3-{\it D}_4\right)\hskip.15in&=
&-\sqrt{8}I_3^{color},  \\
 
\sqrt{\frac{1}{6}}\left({\it D}_3+{\it D}_4-2D_5\right)
&=&-\sqrt{8}Y^{color}.
\eea
\een
Here, $T_{3L}$=diag[1/2, -1/2] for each 
left-handed doublet; $I_3^{color}$=
diag[1/2, -1/2, 0] and $Y^{color}$=\\
$\sqrt{\frac{1}{12}}$diag[1, 1, -2] for each color 
triplet.                                                                                                                                                                                                                                                                                                                                                                                                                                                                                                                                                                                                                                                                                                                                                                                                                                                                                                                                                                                                                                 

\vskip.08in
The mass operators must be block off-diagonal,
color-singlet, and electrically neutral. 
It is easily seen that 
\beq
\gamma_{_1}  \ \ {\rm and}\ \ \gamma_{_6} 
\eeq
in 10 can 
contribute to fermion 
masses if a 10-plet Higgs $H_N$ develops VEV in the first and the sixth 
positions\footnote{The naturalness of developing VEVs at 
two specific positions 
of a 10-plet is a question subject to study[7].}. 
There are four mass operators in 120. They are
\beq
\gamma_{_{1,6}}( i)\gamma_{_2}\gamma_{_{7}}=\gamma_{_{1,6}}D_2;\,\,\,\, 
\gamma_{_{1,6}}(B-L).
\eeq
All non-zero elements of these operator matrices in (17) and (18)
have quantum numbers 
\beq
(SU(3), T_{3L}, T_{3R}, B-L)=(1,\pm 1/2,\mp 1/2, 0).
\eeq
There are four  mass operators in 126 also. Two of them  are
\beq
\gamma_{_{1,6}}(B-L)D_2,
\eeq
which enjoy the same property as described in (19). The other two, from its 
quantum number analysis, are found all to have  zero elements except 
those with  
quantum numbers
$(1,0,\pm 1,\mp 2), \,\,{\rm or}\,\,(1,\pm 1,0,\mp 2)$
and the first one is normally used to give right-handed neutrinos  
huge Majorana masses in order that a ``see-saw" mechanism may take place
to render a tiny left-handed neutrino mass[8]. 
\vskip.08in

A linear combination of the operators in (17) and (18) or (20) can provide 
flexibility to produce desired quark-lepton Dirac mass relations, as
applied in all previous works. Their properties are listed in Table 1, where
subindices $i, \ j$ represent generation (or family) numbers.
\vskip.1in
\begin{center}
{\bf Table} 1 A complete set of $SO(10) $ mass operators\\
\vskip.08in
\begin{tabular}{cccc}\hline\hline \\
operator\hskip.8in&repr.&mass relation\hskip.5in&symm in \\
&&&                                                       family ind.\\
\hline
$\gamma_{_1}$&10&$M^N_{ij}=M^U_{ij}=M^l_{ij}=M^D_{ij}$&S\\
$\gamma_{_6}$&10&$M^N_{ij}=M^U_{ij}=-M^l_{ij}=-M^D_{ij}$&S\\
2$\gamma_{_1}(T_{3R}+T_{3L})$&120&$M^N_{ij}=M^U_{ij}=-M^l_{ij}=-M^D_{ij}$&A\\
2$\gamma_{_6}(T_{3R}+T_{3L})$&120&$M^N_{ij}=M^U_{ij}=M^l_{ij}=M^D_{ij}$&A\\
$\frac{1}{\sqrt{3}}\gamma_{_1}(B-L)$&120&
			$M^N_{ij}=-3M^U_{ij}=M^l_{ij}=-3M^D_{ij}$&A\\
$\frac{1}{\sqrt{3}}\gamma_{_6}(B-L)$&120&
			$M^N_{ij}=-3M^U_{ij}=-M^l_{ij}=3M^D_{ij}$&A\\
$\frac{2}{\sqrt{3}}\gamma_{_1}(T_{3R}+T_{3L})(B-L)$&126&
			$M^N_{ij}=-3M^U_{ij}=-M^l_{ij}=3M^D_{ij}$&S\\
$\frac{2}{\sqrt{3}}\gamma_{_6}(T_{3R}+T_{3L})(B-L)$&126&
			$M^N_{ij}=-3M^U_{ij}=M^l_{ij}=-3M^D_{ij}$&S\\
\hline
\end{tabular}
\end{center}
\vskip.2in

The method used here to produce all the necessary
 matrices on a physical basis
can be used for other $SO$ groups.
\vskip.1in 
{\bf \large  III. Masses and CP Violation}  
\vskip.08in
The most general Yukawa 
term involving one 10 is
\beq
L^Y=\frac{1}{2}\left(g^{ij}_e\bar\Psi_i \gamma_{_N}H_N\Psi_j +
g^{ij}_o\bar\Psi_i
\gamma_{_{11}}\gamma_{_N}H_N \Psi_j\right) +h.c.
\eeq
Note that the expressions
$
(\bar \Psi_i\gamma_{_N}\Psi_j+\bar \Psi_j\gamma_{_N}\Psi_i)\ \ {\rm and}\ \
 (\bar \Psi_ii\gamma_{_{11}}
\gamma_{_N}\Psi_j+\bar \Psi_ji\gamma_{_{11}}
\gamma_{_N}\Psi_i)$ 
are  real.
Both terms in (21) can be CP even if 
Im$g_eg_o^*=0$.
\vskip.08in
According to Eq(21), all fermions may get masses and mix together. 
For the purpose of illustrating how up-down relation goes in $O(10)$
models, we write down  explicitly the 
relevant terms for the third family
 \ben\bea{ccc}
L^Y&=&Re(g_eH_1+g_oH^\prime_6)\bar u_3u_3
+Im(g_eH^\prime_6+g_oH_1)\bar u_3i\bar\gamma_{_5}u_3\\
&&\\
&+&Re(g_eH_1-g_oH^\prime_6)\bar d_3d_3
-Im(g_eH^\prime_6-g_oH_1)\bar d_3i\bar\gamma_{_5}d_3\\
&&\\
&+& {\rm charged\ current\ inter.s} +
{\rm leptonic\ counterpart},
\eea
\een
with $H_6^\prime=-iH_6$ and $g_{e,o}=g_{e,o}^{33}$.
It is easy to check that when $g_e$ and $g_o$ are real, 
the Yukawa term is indeed  CP even if  Re$H_1$ and 
Re$H^\prime_6$ are assigned CP
even while Im$H_1$ and Im$H^\prime_6$ are CP odd. 
The mass relation in (1) is obtained, with a substitution of
$
\langle H_1\rangle=v_1,\ \ \  \langle H^\prime_6\rangle=
\langle -iH_6\rangle=  v_6.
$
\vskip.08in
To discuss the special cases, it is more convenient to use the dual 
representation. This is done by introducing
the following two sets of reduced $\gamma$- matrices, 
which do not belong to 
any Clifford algebra,
\beq
\gamma_{_N}^E=\frac{1}{2}(1+\gamma_{_{11}})\gamma_{_N},\ \ \
\gamma_{_N}^W=\frac{1}{2}(1-\gamma_{_{11}})\gamma_{_N}, \ \ \ 
\gamma^{E\dagger}=\gamma^W.
\eeq
The E and W project operators $(1\pm \gamma_{_{11}})/2$
 are $SO(10)$ invariant. They separate
16 from 16$^*$ in a 32 reduced representation. 
The Yukawa terms expressed in the E-W basis is
\ben
\bea{ccc}
L^Y&=&\frac{1}{2}\left(G^{ij}_E\bar\Psi_i\gamma^E_{_N}H_{N}\Psi_j+
G^{ij}_W\bar\Psi_i\gamma^W_{_N}H_N\Psi_j\right)+h.c.\hskip1.7in\\
&&\\
&=&G^{33}_E\left(H_{u}\bar u_{3L}u_{3R}+H_{d}\bar d_{3L}d_{3R}\right)
+G^{33}_W\left(H_{d}\bar u_{3L}u_{3R}+H_{u}\bar d_{3L}d_{3R}\right) +h.c.\\
&&\\
&+&   \ leptonic\ and\ charged+ 
\cdot\cdot\cdot, \hskip2.3in
\eea
\een
where 
$
H_{u,d}=H_1\pm iH_{6},\ \ \  G_{E,W}=g_e\pm g_o.
$
$H_u$ and $H_d$ are respectively in 5 and 5$^*$ of $SU(5)$ which are 
respectively up and down
components of 10 (please compare with Eq (7)).   
\vskip.08in

 Let us now return to the special cases discussed in Section I.\\
 {\bf Case 1)}, $v_d=\langle H_d\rangle=0$: One obtains
\beq
m_t/m_b=(g_e+g_o)/(g_e-g_o)=G_E/G_W.
\eeq
It can give the phenomenological mass ratio with only one VEV. But if 
Re$g_eg^*_o=0$  (which corresponds to a maximal 
CP phase when $g_eg*_o\ne 0$) 
one is forced to have  the trivial $O(10)$ mass relation $|m_t|=|m_b|$. 
\vskip.08in
The mass matrix for the up-type quarks and down-type quarks are, in the case 
of three families of quarks and leptons,
\beq
M^U=G_E^{ij}v, \ \ \ M^D=G_W^{ij}v,\ \ \  (i,\ j= 1,\ 2,\ 3)
\eeq
while the Yukawa couplings for the $H_d$ field, which does not develop VEV, 
are 
\beq
Y^U=G_W^{ij}, \ \ \ Y^D=G_E^{ij}.
\eeq
Although all fermions obtain their masses from the one
and only vacuum expectation value $v=v_u$, there
 does not necessarily exist a proportionality relation 
between the masses and the
Yukawa coupling constants of the $H_d$ field, unless $g_o=0$ or $g_e=0$.
Therefore, the $H_d$  mediated FCNC exist in general, as does new CP 
violation in the Yukawa sector, if Im$g_eg_o^*\ne 0$. When $g_o=0$, half of 
the Higgs degrees of freedom decouple from the Yukawa sector. \\
\noindent {\bf Case 2)},  $G_W=0$: 
 $H_{u}$ only provides up-type masses; and $H_d$ only provides 
down-type masses. This case has attracted the most 
attention in the literature.
But it raises the question of how  one can forbid the $G_W$ term by certain
symmetries. 
The only candidate within the $O(10)$
group seems to be the 
$O(10)$ dual transformation
$
\gamma_{_N}\rightarrow \gamma_{_{11}}\gamma_{_N}.
$
The E-type term is self-dual and the W-type term is anti-self-dual.
An extra $U(1)$ global symmetry appears in the Yukawa sector, along with 
self-duality. 
\vskip.08in

Another representation can separate  $H_{\pm}$  at the beginning,
where 
$
H_{I\pm}=H_I\pm iH_{I+5}.\ \ \ (I=1 \ {\rm to}\ 5)
$
$H_{I+}$ and $H_{I-}$ are in 5$_1$ and 5$^*_2$ respectively of $SU(5)$ and 
$10=(5_1, 5_2^*)$ in this representation.  
The associated  reduced $\gamma $- matrices are
$
\gamma_{_{I\pm}}=(\gamma_{_I}\pm i\gamma_{_{I+5}})/2. 
\ \ \ (I=1 \ {\rm to}\ 5) 
$
The P even ($G_E=G_W$) Yukawa interaction is then
\beq
\bar\Psi\sum_{I=1}^5(\gamma_{_{I-}}H_{I+}+\gamma_{_{I+}}H_{I-})\Psi.
\eeq
This representation is convenient for discussions of 
charged currents and gauge interactions.
\vskip.2in
{\bf\large IV. Discussions }
\vskip.08in
It has been explicitly shown that a general $O(10)$ grand unification
 model with one 10-plet Higgs provides a natural motivation for the most 
general two Higgs doublet model (2HDM) at low energies as discussed in 
detail in Ref. 3. In a general MOTM, there must be FCNC,
if trivial up-down $O(10)$ mass relations is to be avoided.
  It is easy to see that this behavior also appears
 when one Higgs 120 or 126 contributes to Dirac fermion masses,
except that 120 contributes only inter-family masses. 
\vskip.08in

In addition to  self-duality,
 one can also rule out the W- current- H coupling by the use of:
a) supersymmetry; b) an extra $U(1)$ quantum
number; c) a discrete symmetry of an
order higher than five; d) a complex nonabelian group. 
\vskip.08in

Actually, certain amount of FCNC is tolerable within  the accuracy of
the present experimental data, 
as discussed in Ref. 3. While self-duality can make
$m_b\ne m_t$,
the most general condition for $m_b\ne m_t$ is 
$$\bea{cccccccc}
&{\rm Re}(g_eg_o^*)&\ne &0,& \hskip.3in &{\rm Re}(v_1v_6^*)&\ne &0,\\
{\rm or}& |G_W|&\ne &|G_E|,&\hskip.3in &|v_u|&\ne &|v_d|.
\eea$$
When FCNC is allowed there are 
 possibilities to realize the desired up-down mass ratio by a 
combination of two VEVs and two coupling constants. 
\vskip.08in

It is very interesting that within the realm of the explicit CP invariant 
MOTM, one can adjust the up-down mass ratio in one 
family by adjusting $g_o/g_e$
and $v_u/v_d$.
Therefore it is possible to find
an  $O(10)$ model with a CP invariant Lagrangian. 
 In such a model, CP violation all will be spontaneous.
\vskip.08in 

In conclusion, there is a correlation between explicit CP violation and
fermion mass relation in an $O(10)$ grand unification model. 
The $O(10)$ Higgs
multiplets which may contribute to fermion masses have two or more 
 doublets whose CP transformation properties are different. 
The introduction of self-duality, or symmetries beyond the $O(10)$ group, is 
crucial for getting rid of flavor changing neutral currents if they are
undesired. 
For convenience of this study, a new form of explicit $O(10)$ gamma matrices 
are given, based on physical representations of spinors and vectors.  
\vskip.08in
Except for the $O(10)$ group, other $O(2n)$  
($n\ge 2$) and $E_6$ models may also have similar  correlation
between CP violation and masses. 
\vskip.08in
One of the authors (DD) sincerely thanks H.J. He, 
for very useful discussions. Comments from R. Arnowitt is appreciated.
The CERN theory 
group and the DESY theory group are acknowledged for their 
hospitality during DD's visit.
The work of YL is supported in part by the
U.S. Department of Energy under contract DOE/ER/01545-675. 
The work of DD is supported in part by the U.S. Department of Energy under 
contract DE-FG03-95 ER40914/A00. 
 
\clearpage
{\bf References}
\begin{itemize}
\begin{enumerate}

\item H. Georgi, Proc. AIP, Ed. C.E. Carlson, Meeting at William \& Mary 
College, 1974; H. Fritzsch and P. Minkowsky, Ann. Phys. (NY) 93 (1975) 193;
 M. Gell-Mann, P. Ramond, and R. Slansky, Rev. Mod. Phys. 50 (1978) 721.
\item R. Peccei and H. Quinn, Phys. Rev. D16 (1979) 1977.
\item For a recent discussion of two Higgs doublet models, see, Y.L. Wu 
and L. Wolfeinstein, Phys. Rev. Lett. 73 (1995) 1762 and therein. 
See also, L. Z. Sun, and Y. Y. Liu, Phys. Rev. D53 (1996) 2411; 
D. Atwood, L. Reina and A. Soni, hep-ph/960321;
L.J. Hall and S. Weinberg, Phys. Rev. D48 (1993)  979.
\item For  recent studies of  $O(10)$ models, see e.g., 
 K.C. Chou and Y.L. Wu,  Phys. Rev. D {\bf 53} (1996) 
(Rapid Communication);
C.H. Albright and S. 
Nandi, FERMILAB-PUB-95/107-T and HEP-PH/9505383, N. Haba, C. Hattori, 
M. Matsuda, and T. Matsuoka, HEP-PH/9512 and references therein.
\item A recent discussion of physics beyond the CKM 
matrix can  be found in, e.g.  D.D.
 Wu, PVAMU-HEP-8-95, to appear in Phys. Lett. B, and references therein.
\item See,  e.g. S. Rajpoot, and P. Sithikong, Phys. Rev. D22 (1980) 2244;
   F. Wilczek and A. Zee, Phys. Rev. D25 (1982) 553; J.A. Harvey, D. B. 
Reiss, and P. Remond, Nucl. Phys. B199 (1982) 223.                      
\item H. Ruegg, Phys. Rev. D22 (1980) 2040; D.D. Wu, Nucl. Phys. 
199B (1981) 523.
\item M. Gell-Mann, P. Ramond, and R. Slansky, in {\it Supergravity}, ed.
D. Freedman and P. van Niuenhuizen, (North Holland, 1979); 
T. Yanagida, KEK Proceedings
(1979); R. N. Mohapatra and G. Senjanovic, Phys. Rev. Lett. 44 (1980) 912.

\end{enumerate}
\end{itemize}
\end{document}